\documentclass[twocolumn,english]{revtex4}
\usepackage[T1]{fontenc}
\usepackage[latin1]{inputenc}
\usepackage{graphicx}

\makeatletter

\providecommand{\tabularnewline}{\\}

\usepackage{babel}
\makeatother
\begin{document}

\title{Action Potential Onset Dynamics and the Response Speed of Neuronal
Populations}

\author{B.~Naundorf, T.~Geisel and F.~Wolf}

\affiliation{Max-Planck-Institut für Strömungsforschung and Fakultät für Physik,
Universität Göttingen, 37073 Göttingen, Germany}

\email{bjoern@chaos.gwdg.de}

\date{\today}

\begin{abstract}
The result of computational operations performed at the single cell
level are coded into sequences of action potentials (APs). In the
cerebral cortex, due to its columnar organization, large number of
neurons are involved in any individual processing task. It is therefore
important to understand how the properties of coding at the level
of neuronal populations are determined by the dynamics of single neuron
AP generation. Here we analyze how the AP generating mechanism determines
the speed with which an ensemble of neurons can represent transient
stochastic input signals. We analyze a generalization of the $\theta$-neuron,
the normal form of the dynamics of Type-I excitable membranes. Using
a novel sparse matrix representation of the Fokker-Planck equation,
which describes the ensemble dynamics, we calculate the transmission
functions for small modulations of the mean current and noise noise
amplitude. In the high-frequency limit the transmission function decays
as $\omega^{-\gamma}$, where $\gamma$ surprisingly depends on the
phase $\theta_{s}$ at which APs are emitted. If at $\theta_{s}$
the dynamics is insensitive to external inputs, the transmission function
decays as (i) $\omega^{-3}$ for the case of a modulation of a white
noise input and as (ii) $\omega^{-2}$ for a modulation of the mean
input current in the presence of a correlated and uncorrelated noise
as well as (iii) in the case of a modulated amplitude of a correlated
noise input. If the insensitivity condition is lifted, the transmission
function always decays as $\omega^{-1}$, as in conductance based
neuron models. In a physiologically plausible regime up to $1\textrm{kHz}$ the
typical response speed is, however, independent of the high-frequency
limit and is set by the rapidness of the AP onset, as revealed by
the full transmission function. In this regime modulations of the
noise amplitude can be transmitted faithfully up to much higher frequencies
than modulations in the mean input current. We finally show that the
linear response approach used is valid for a large regime of stimulus
amplitudes.
\end{abstract}
\maketitle

\section{Introduction}

Neurons are the basic building blocks of neural networks and thus
constitute the computational units of the brain. They dynamically
transform synaptic inputs into output action potential (AP) sequences.
To conceive the complex computational capabilities of the brain, it
is crucial to understand this transformation and to identify simple
neuron models which accurately reproduce the dynamical features of
cortical neurons.

Here we study this mapping in a reduced neuron model. This model is
obtained by a generalization of the $\theta$-neuron \cite{Ermentrout84,Gutkin98},
which is a canonical phase oscillator model of excitable neuronal
membranes exhibiting Type-I excitability. Phase oscillator models
have a long history in physics and biology \cite{Coddington55,Winfree74,Guckenheimer75,Winfree01}
and recently they were introduced in theoretical neuroscience \cite{Ermentrout84}.
In contrast to integrate-and-fire models, which are phenomenological
models of cortical neurons, they can be derived from the limit cycle
dynamics of conductance based neuron models, reducing the complex
dynamics which usually incorporates many degrees of freedom to a single
phase variable. This reduction is an important prerequisite for analytical
studies of either the dynamics of single neurons or of neural networks.

Cortical neurons \emph{in vivo} are subject to an immense synaptic
bombardment, resulting in large fluctuations of their membrane potential
(MP) \cite{Destexhe99,Volgushev00,Anderson00} and irregular action
potential firing \cite{Softky93}. Because the exact computational
role of these fluctuations is largely unknown, it was suggested to
treat them as a random process, dividing the synaptic input into a
mean input current and a random fluctuating contribution \cite{Tuckwell88}
with a given correlation time $\tau_{c}$. The fluctuations can serve
as a potentially independent information channel because when the
afferent activity of a neuron changes, not only the mean input is
affected, but also the amplitude of the fluctuations\cite{Vreeswijk96,Brunel99,Lindner02}.

The stationary response properties of the classical $\theta$-neuron
subject to fluctuating input currents were calculated in \cite{Gutkin98,Lindner03}
for a temporally uncorrelated input and in \cite{Brunel03} for a
temporally correlated input current. Both studies showed that the
$\theta$-neuron can reproduce the stationary response properties
exhibited by many cortical neurons, i.e.~a square-root dependence
of the firing rate on the input current close to threshold for small
noise amplitudes \cite{Tateno04} and irregular firing in the noise
driven regime. Despite its success to reproduce the stationary firing
behavior of cortical neurons, the $\theta$-neuron lacks a crucial
dynamical feature: The fast action potential upstroke exhibited by
conductance based neuron models. Here we study a generalization of
the classical $\theta$-neuron with an adjustable action potential
onset speed, introducing a phenomenological term which mimics the
fast activation of sodium channels. 

We derive the time dependent response in the presence of temporally
correlated noise to a modulation in the mean input current and a modulation
in the noise amplitude. For both modulation paradigms we calculate
the high frequency limit. In this limit, the response amplitude decays
as $\omega^{-\gamma}$, where the integer exponent $\gamma$ is completely
independent of the action potential onset dynamics and surprisingly
only depends on the oscillator phase $\theta_{s}$, at which an action
potential is emitted: If at $\theta_{s}$ the dynamics is insensitive
to external inputs, the transmission function decays as (i) $\omega^{-3}$
for the case of a modulation of an uncorrelated noise amplitude and
as (ii) $\omega^{-2}$ for a modulation of the mean input current
in the presence of a correlated and uncorrelated noise as well as
(iii) in the case of a modulated amplitude of a correlated noise input.
If the insensitivity condition is lifted, the transmission function
always decays as $\omega^{-1}$, as in conductance based neuron models.

The full transmission function is then calculated via the eigenvalues
and eigenfunctions of the Fokker-Planck operator, which describes
the dynamics of the ensemble averaged probability density function.
The eigenvalues and eigenfunctions are computed using a high performance
iterative scheme, the Arnoldi method \cite{Trefethen97,Lehoucq98},
from a novel sparse matrix representation of the Fokker-Planck operator.
This method allows for a fast computation and high numerical precision,
hard to achieve by direct numerical simulations.

We then demonstrate that the response amplitudes for the classical
$\theta$-neuron typically exhibit a cut-off behavior, where the cut-off
frequency, which is closely linked to the spectral properties of the
Fokker-Planck operator, is approximately given by the neurons stationary
firing rate. Stimulations at frequencies larger than the cut-off frequency
are strongly damped. For an increasing action potential onset speed
at a fixed stationary rate, stimuli with much larger frequencies can
be transmitted almost unattenuated. We show that the response amplitude
for the case of a noise modulation typically decays much slower than
in the case of a mean input current modulation. 

The impact of noise on the dynamic response properties was previously
almost exclusively studied in integrate-and-fire models \cite{Lapicque07,Tuckwell88}.
The first studies were pioneered by Knight \cite{Knight72}, who considered
a simple integrator model, in which the firing threshold is drawn
randomly, every time an action potential occurs. These results were
then extended to models, where the reset voltage was also drawn randomly,
and to models in which the the input changed either very slowly, or
to spike response models, where the input is assumed to change very
fast \cite{Gerstner00}. Recently, the impact of current noise on
the dynamical response of the leaky integrate-and-fire model was investigated
\cite{Brunel99,Brunel00,Brunel01,Fourcaud02,Lindner02}. In these
studies it was shown that integrate-and-fire models driven by a synaptic
fluctuating input exhibit a linear response amplitude which does not
decay to zero in the high frequency limit. This lead some to the conclusion
that cortical neurons can transmit information instantaneously \cite{Brunel01,Lindner02}.
Only recently, this interpretation was questioned by two studies \cite{Naundorf03,Fourcaud03}
which demonstrated that the unattenuated transmission of high frequency
signals in integrate-and-fire models are more a consequence of the
oversimplification of the model rather than property of real neurons.

\section{Material and Methods}

\subsection{Model}

\begin{figure}
\includegraphics[%
  width=1.0\columnwidth]{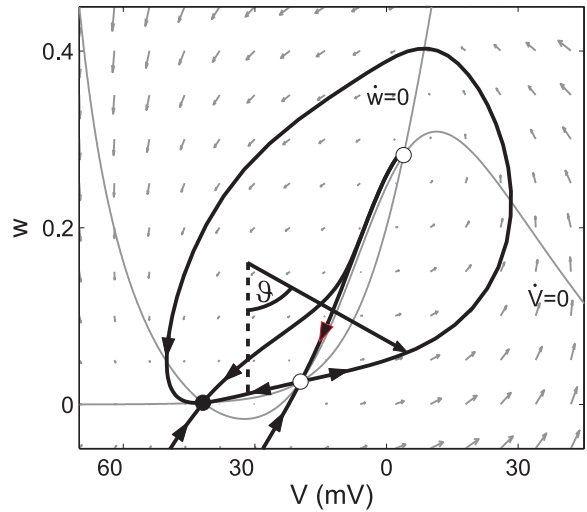}

\caption{\label{cap:MPPhaseSpace}Phase plane of a type-I single compartment
conductance based model (Morris-Lecar model \cite{Morris81}) in the
excitable regime (filled dot: stable fixed point, open dots: unstable
fixed points). Gray lines are the nullclines, denoted by $\dot{w}=0$
and $\dot{V}=0$. Black lines are stable and unstable manifolds of
the saddle and the node. The excitable dynamics can be reduced to
a phase oscillator with one degree of freedom parameterized by the
angle $\vartheta$.}
\end{figure}
The model is based on the normal form of the dynamics of type-I membranes
at the bifurcation to repetitive firing. Conductance based neuron
models which exhibit Type-I excitability typically undergo a saddle-node
bifurcation of codimension one, when brought to repetitive firing.
A center-manifold reduction at the bifurcation point leads to the
following normal form \cite{Strogatz95}:\begin{equation}
C\dot{\bar{V}}=\frac{g}{V_{0}}\left(\bar{V}-V^{*}\right)^{2}+\left(\bar{I}(t)-I_{c}\right),\end{equation}
which is a dynamical equation for the MP  $\bar{V}$ of the neuron.
The input current relative to the rheobase $I_{c}$ of the neuron
is denoted by $\bar{I}(t)$. The constants $A$ and $V^{*}$ can be
deduced from a given multidimensional conductance based model. It
is convenient to introduce dimensionless quantities $V$and $I$:\begin{eqnarray}
V & = & \left(\bar{V}-V^{*}\right)/V_{0}\\
I(t) & = & \left(\bar{I}(t)-I_{c}\right)/\left(gV_{0}\right)\end{eqnarray}
and the effective time constant:\begin{equation}
\tau=C/g\end{equation}
The rescaled dynamics is then given by:\begin{equation}
\tau\dot{V}=V^{2}+I(t)\label{eq:NormalForm}\end{equation}
For $I(t)>0$, the MP has a finite {}``blow-up'' time, meaning that
it needs a finite time to get from $-\infty$ to $+\infty$, where
both ends of the real axis are identified, turning the model into
a phase oscillator. The normal form Eq.~(\ref{eq:NormalForm}) is
equivalent to a phase oscillator, the $\theta$-neuron \cite{Ermentrout84,Gutkin98}.
Its equation of motion,\begin{equation}
\tau\dot{\theta}=\left(1-\cos\theta\right)+I(t)\left(1+\cos\theta\right)\label{eq:ThetaNeuron}\end{equation}
is found by substituting $V=\tan\left(\theta/2\right)$ with the angle
variable $\theta$ in the interval $(-\pi,\pi]$. 

In the model, a spike is emitted each time $\theta$ reaches the value
$\theta_{s}$. By choosing $\theta_{s}=\pi$, the original $\theta$-neuron
is obtained. Figure \ref{cap:MPPhaseSpace} illustrates schematically
the reduction of a conductance based neuron model to a phase oscillator
model.

Although the $\theta$-neuron is the normal form of the dynamics at
the bifurcation, it lacks the rapid AP onset exhibited by conductance
based neuron models and real neurons. To account for this dynamical
feature we generalized the model to reflect the rapid depolarization
of the membrane resulting from the fast kinetics of sodium conductances
in the following way:\begin{equation}
\tau\dot{V}=V^{2}+I(t)+\alpha\left(1+\tanh(\beta V)\right),\label{eq:GeneralizedNormalForm}\end{equation}
where we introduced two additional parameters $\alpha$ and $\beta$.
The sigmoidal term phenomenologically models the part of the sodium
channel activation curve, which is not included in the $V^{2}$-term
of the normal form. The parameter $\alpha$ controls the sodium peak
conductance and the parameter $\beta$ the width of this activation
curve. Both parameters control the rapidness of the AP onset. As for
the normal form an equivalent phase oscillator equation can be found
by substituting $V=\tan\left(\theta/2\right)$:\begin{eqnarray}
\tau\dot{\theta} & = & \left(1-\cos\theta\right)+\left(1+\cos\theta\right)\cdot\label{eq:GeneralizedThetaNeuron}\\
 &  & +\left\{ I(t)+\alpha\left(1+\tanh(\beta\tan\left(\theta/2\right)\right)\right\} \nonumber \end{eqnarray}

\subsection{Fluctuating input currents}

\emph{In vivo}, neurons are subject to an ongoing synaptic bombardment,
resulting in a fluctuating MP. To model this situation, we assume
a temporally fluctuating input current,\begin{equation}
I(t)=I_{0}+\sigma z(t),\label{eq:CorrelatedInputCurrent}\end{equation}
composed of a mean $I_{0}$ and a stationary fluctuating part $\sigma z(t)$,
where $z(t)$ is an Ornstein-Uhlenbeck process with a given correlation
function $\left\langle z(t)z(t')\right\rangle =\exp\left(-t/\tau_{c}\right)$.
Thus $z(t)$ obeys the Langevin equation \cite{Gardiner85},\begin{equation}
\tau_{c}\frac{d}{dt}z(t)=-z+\sqrt{\tau}\eta(t)\label{eq:DeqNoiseInput}\end{equation}
with $\left\langle \eta(t)\right\rangle =0$ and $\left\langle \eta(t)\eta(t')\right\rangle =\delta(t-t')$.
Eq.~(\ref{eq:ThetaNeuron}) and Eq.~(\ref{eq:GeneralizedThetaNeuron})
describe a realization of the dynamics of a single neuron. Since the
input is fluctuating and we are interested in coding at population
level it is natural to consider an ensemble of such units, described
by the time dependent probability density function $P(\theta,z,t)$.
Its dynamics is determined by the Fokker-Planck equation \cite{Risken96}:
\begin{equation}
\partial_{t}P(\theta,z,t)=\hat{L}P(\theta,z,t),\label{eq:FokkerPlanckCorrelated}\end{equation}
with, \begin{eqnarray}
\hat{L} & = & -\tau^{-1}\partial_{\theta}\left\{ \left(1-\cos\theta\right)\right.\nonumber \\
 &  & +\left(I_{0}+\sigma z+\alpha\left(1+\tanh\left(\beta\tan\left(\theta/2\right)\right)\right)\right)\nonumber \\
 &  & \left.\vphantom{\left(I_{0}\right)}\cdot\left(1+\cos\theta\right)\right\} +\tau_{c}^{-1}\partial_{z}\left(z+\frac{\tau}{2\tau_{c}}\partial_{z}\right).\end{eqnarray}
The boundary conditions for $P(\theta,z,t)$ are periodic in the $\theta$-
and natural in the $z$-direction.

\subsection{Time dependent firing rate}

The ensemble averaged firing rate is given by the probability current
across the line $\theta=\theta_{s}$ with positive velocity. At $\theta_{s}=\pi$
the dynamics is independent of the input current $I(t)$ and the rate
is equal to the probability current through the entire line $\theta=\pi$:\begin{equation}
\nu(t)=2\int_{-\infty}^{\infty}dz\, P(\pi,z,t)\label{eq:StationaryRateCorr}\end{equation}
Although quite convenient for analytical considerations, the definition
of this spike-phase is, however, rather arbitrary. In the normal form,
the point $\theta_{s}=\pi$ corresponds to the point $V=\infty$,
where the model reflects least the dynamics at the bifurcation. To
assess if this particular choice has any influence on the dynamical
response properties of the model, we also calculate the firing rate
at $\theta_{s}=\pi-\delta$. The probability current through this
line is given by:\begin{eqnarray}
J_{\theta} & = & \tau^{-1}\int_{-\infty}^{\infty}P(\theta_{s},z,t)\left((1-\cos\theta_{s})\right.\label{eq:CurrentWithFiniteDelta}\\
 &  & \!\!\!\left.+(I_{0}+\sigma z+\alpha(1+\tanh(\beta\tan(\theta_{s}/2)))\right)\, dz\nonumber \end{eqnarray}
The rate is, however, not exactly given by the flux $J_{\theta}$.
There is a contribution from trajectories, which are driven back below
the threshold due to the external fluctuations. For a correlated input
current, however, the introduced error is exponentially small. This
can be seen in Eq.~(\ref{eq:GeneralizedThetaNeuron}). For small
values of $\delta$, the probability distribution $P(\theta,\dot{\theta})$
around $\pi-\delta$ is a Gaussian with a mean value $2-\delta^{2}$
and a width $\propto\delta^{2}$. The negative part of this Gaussian
is proportional to: \begin{equation}
\left(2\pi\delta^{4}\sigma^{2}\right)^{-1/2}\int_{-\infty}^{0}\exp\left(-\frac{\left(x-(2-\delta^{2})\right)^{2}}{2\delta^{4}\sigma^{2}}\right)\, dx\end{equation}
For all practical purposes ($\delta<0.5$ and $\sigma<1$), this integral
is smaller than $10^{-10}$. We will see, however, that the definition
$\theta_{s}=\pi$ in the classical $\theta$-neuron qualitatively
changes the dynamic response of the model in the high frequency limit.

\subsection{Parameter choice}

Before discussing the stationary and dynamical properties of the generalized
$\theta$-neuron we would like to define a biologically plausible
parameter regime. The parameters which we need to fix are the time
constant $\tau$, the mean input current $I_{0}$, the strength of
the fluctuating input $\sigma$ and the synaptic input correlation
time $\tau_{c}$. An estimate of the correlation time of the MP is
given by approximating the dynamics for $I_{0}<0$ near the stable
fixed point by an Ornstein-Uhlenbeck process. Straightforward linearization
around the stable fixed point at $-\sqrt{I_{0}}$ then yields:\begin{equation}
\tau_{\mathrm{relax}}\approx\tau\left(2\sqrt{I_{0}}\right)^{-1}\label{eq:RelaxTimeMP}\end{equation}
In the subthreshold noise-driven regime, which we will discuss in
the following, we choose $I_{0}=-0.1$. The time constant $\tau$
is then adapted via Eq.~(\ref{eq:RelaxTimeMP}), to achieve a relaxation
time of approximately $5\textrm{ms}$, which leads to values for $\tau$
of approximately $3\textrm{ms}$. 

The parameters $\alpha$ and $\beta$ parameterize the sodium activation
curve, which, in conductance based models, determines the speed at
the action potential onset. For the following numerical treatment
we keep $\beta$, which mediates the width of the activation curve
and is an intrinsic physiological parameter, fixed to a value of 20.
The parameter $\alpha$, which represents the sodium peak conductance,
is changed in the range from $0$ to $1$.

\begin{figure}
\includegraphics[%
  width=1.0\columnwidth]{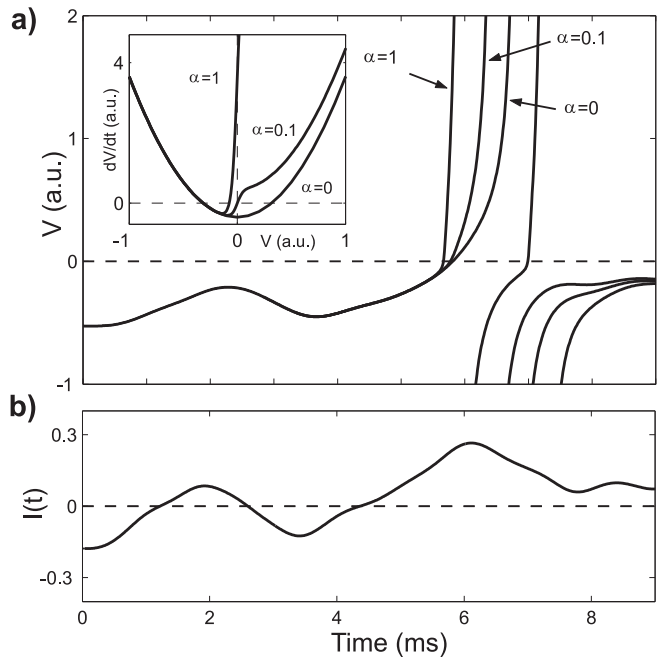}

\caption{\label{cap:Sample-MP-Traj}Increasing $\alpha$ leads to a sharper
action potential onset. (a) Sample MP trajectories for $\alpha=0$,
$\alpha=0.1$ and $\alpha=1$. The inset shows the deterministic part
of Eq.~(\ref{eq:GeneralizedNormalForm}). (b) Fluctuating input current
$I(t)$. Parameters: $\tau_{c}=1.5\textrm{ms}$, $\sigma=0.3$, $I_{0}=-0.1$
and $\beta=20$. Right before AP onsets the trajectories are virtually
identical. }
\end{figure}
Figure \ref{cap:Sample-MP-Traj} shows three sample realizations of
Eqs.~(\ref{eq:GeneralizedThetaNeuron}, \ref{eq:CorrelatedInputCurrent})
for different values of the parameter $\alpha$. If the input current
is positive for a sufficient amount of time, action potentials are
initiated. With increasing values of $\alpha$ the sharpness at the
onset increases, while the subthreshold fluctuations are not affected.

\subsection{Dynamic Response Theory}

\begin{figure*}
\includegraphics[%
  width=0.90\textwidth]{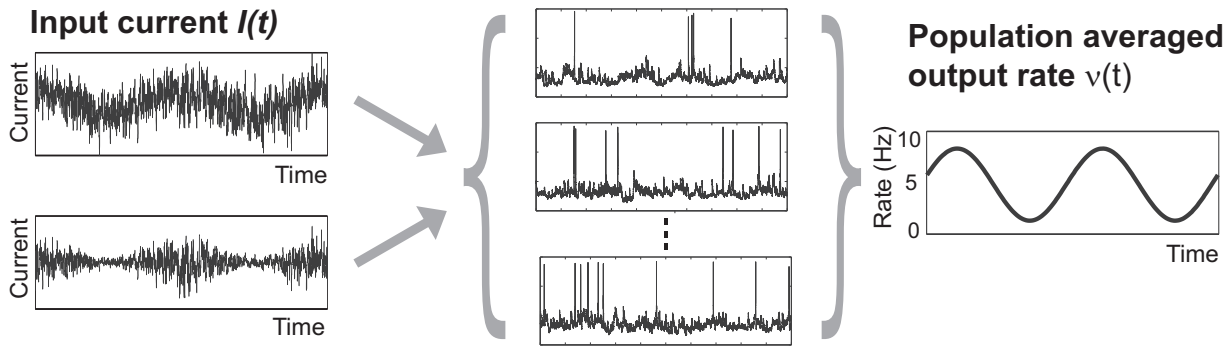}

\caption{\label{cap:MDynResMethod}Sketch of the population response paradigm.
An ensemble of neurons receives a modulated noisy input current or
a current, where the noise amplitude is modulated. The noise realization
which each neuron receives is different, leading to different MP traces
and AP sequences. The output quantity is the population averaged firing
rate in the interval $[t,t+dt)$, $\nu(t)$. }
\end{figure*}
For time-dependent input currents $\varepsilon I(t)$, the Fokker-Planck
operator $\hat{L}(\theta,z,t)$ can always be split into two parts:\begin{equation}
\hat{L}(\theta,z,t)=\hat{L}_{0}(\theta,z)+\varepsilon\hat{L}_{1}(\theta,z,t),\end{equation}
where $\hat{L}_{0}(\theta,z)$ is the time-independent part and $\hat{L}_{1}(\theta,z,t)$
contains all time-dependencies of the external input. In the following
we require that the time-dependent inputs are small in magnitude,
i.e.~$\varepsilon\ll1$. We then expand the general time-dependent
solution in powers of $\varepsilon:$\begin{equation}
P_{\mathrm{TD}}(\theta,z,t)=P_{0}(\theta,z)+\varepsilon\tilde{P}(\theta,z,t)+\mathcal{O}(\varepsilon^{2})\end{equation}
Inserting this solution into the Fokker-Planck equation and keeping
only terms up to linear order in $\varepsilon$ leads to a dynamical
equation for the time dependent part of the density $\tilde{P}(\theta,z,t)$:\begin{equation}
\partial_{t}\tilde{P}(\theta,z,t)=\hat{L}_{0}(\theta,z)\tilde{P}(\theta,z,t)+\hat{L}_{1}(\theta,z,t)P_{0}(\theta,z)\label{eq:LinResFPG}\end{equation}
Formally the solution of this equation is given by:\begin{equation}
\tilde{P}(\theta,z,t)=\int_{-\infty}^{t}e^{\hat{L}_{0}(t-t')}\hat{L}_{1}(\theta,z,t)P_{0}(\theta,z)\, dt'\label{eq:LinearResponseIntegral}\end{equation}
In the following we will consider stimuli of the type:\begin{equation}
\hat{L}_{1}(\theta,z,t)=e^{i\omega t}\hat{L}_{1}(\theta,z)\end{equation}
 Eq.~(\ref{eq:LinearResponseIntegral}) can then be readily solved,
yielding:\begin{equation}
\tilde{P}(\theta,z,t)=\sum_{k}\frac{c_{k}}{i\omega-\lambda_{k}}P_{k}(\theta,z)e^{i\omega t}\label{eq:LinResponseSolEF}\end{equation}
The $c_{k}$ are the expansion coefficients of $\hat{L}_{1}(\theta,z)P_{0}(\theta,z)$
into the eigenfunctions $P_{k}(\theta,z)$ of $\hat{L}_{0}(\theta,z)$.
The time-dependent firing rate is given by Eq.~(\ref{eq:StationaryRateCorr}):\begin{eqnarray}
\nu(t) & = & \tau^{-1}\int_{-\infty}^{\infty}dz\,\left((1-\cos\theta_{s})\right.\nonumber \\
 &  & \left.+\left(I_{0}+\sigma z+\alpha(1+\tanh(\beta\tan(\theta_{s}/2)))\right)\right)\nonumber \\
 &  & \cdot\left(P_{0}(\theta_{s},z)+\varepsilon\tilde{P}(\theta_{s},z,t)\right)\nonumber \\
 & =: & \nu_{0}+\varepsilon\nu_{1}(\omega)e^{i(\omega t+\varphi(\omega))}\label{eq:LROutputRate}\end{eqnarray}
In the following we will consider two types of external stimulations: 

\begin{enumerate}
\item Modulations in the mean input current: \[
I_{0}\longrightarrow I_{0}+\varepsilon e^{i\omega t}\]

\item Modulations in the noise amplitude: \[
\sigma\longrightarrow\sigma+\varepsilon e^{i\omega t}\]

\end{enumerate}

\subsection{\label{sub:High-frequency-limit}High frequency limit}

In this section we sketch how to analytically calculate the asymptotic
decay of $\nu_{1}(\omega)$ in the limit $\omega\to\infty$. Inserting
Eq\@.~(\ref{eq:LinResponseSolEF}) into Eq.~(\ref{eq:LinResFPG})
leads to:\begin{equation}
\left(i\omega-\hat{L}_{0}\right)\tilde{P}(\theta,z,t)e^{-i\omega t}=\hat{L}_{1}P_{0}(\theta,z)\label{eq:FirstOrderHighFreqLimit}\end{equation}
If the right hand side vanishes at $\theta=\theta_{s}$, $\tilde{P}(\theta,z,t)$
has to decay at least as $\omega^{-2}$. Differentiation of Eq.~(\ref{eq:LinResFPG})
with respect to $t$ and subsequent reinsertion leads to:\begin{equation}
\left(\omega^{2}+\hat{L}_{0}\right)\tilde{P}(\theta,z,t)=-\hat{L}_{0}\hat{L}_{1}P_{0}(\theta,z)\label{eq:MethodSecOrderHighFreqLimit}\end{equation}
If now the right hand side vanishes at $\theta=\theta_{s}$, Eq.~(\ref{eq:LinResFPG})
has to be differentiated again, until, after reinsertion, the right
hand side is different from zero.

\subsection{Matrix Method}

As demonstrated the dynamical response properties of the generalized
$\theta$-neuron to small time-dependent inputs are completely determined
by the spectrum and eigenfunctions of the Fokker-Planck operator $\hat{L}$.
To compute the dynamical response properties in the presence of a
temporally correlated noise current for arbitrary stimulation frequencies
we expand $\hat{L}$ into a complete orthonormal basis leading to
a sparse matrix representation for which we compute the eigenvalues
and eigenfunctions numerically. This approach has the advantage that
the response properties can be computed with very high accuracy. The
two subtleties we will have to deal with are that (1) the resulting
matrix is very large in the parameter regime we are interested in
(up to $10^{6}\times10^{6}$) and (2) the operator $\hat{L}$ is not
Hermitian and thus standard diagonalization procedures such as the
Lanczos algorithm can not be applied. We solved both problems by using
a basis-set, which results in a very sparse matrix representation,
and by using a high performance iterative scheme, the Arnoldi method
\cite{Lehoucq98}, to compute the eigenfunctions and the spectrum
of this matrix to a high numerical accuracy.

\subsection{\label{sub:MatrixMethodCorrelated}Eigenvalues and eigenfunctions
for a correlated noise input}

\subsubsection{Matrix equation}

We first replace the probability density $P(\theta,z,t)$ in an eigenmode
Ansatz with $e^{\lambda_{k}t}P_{k}(\theta,z)$. Inserting this into
Eq.~(\ref{eq:FokkerPlanckCorrelated}) the exponential cancels:\begin{equation}
\lambda_{k}P_{k}(\theta,z)=\hat{L}P_{k}(\theta,z)\label{eq:EigenvalueEquationL}\end{equation}
Due to the imposed boundary conditions, the set $\left\{ \lambda_{k}\right\} $,
i.e. the spectrum of $\hat{L}(\theta,z)$ is discrete. There is, however,
a macroscopic drift in the system, meaning that detailed balance is
not fulfilled and thus $\hat{L}$ is not Hermitian \cite{Gardiner85}.
This means that the resulting spectrum $\left\{ \lambda_{k}\right\} $
and the corresponding eigenfunctions $P_{k}(\theta,z)$ are complex.
By complex conjugation of Eq.~(\ref{eq:EigenvalueEquationL}) it
is easy to show that to every eigenvalue $\lambda_{k}$ with the corresponding
eigenfunction $P_{k}(\theta,z)$, an eigenvalue $\lambda_{k}^{*}$
with the eigenfunction $P_{k}^{*}(\theta,z)$ exists. This guarantees
that a real solution can always be constructed. The solution with
$\lambda_{0}=0$ corresponds to the stationary density and the time
dependent solution can always be given in terms of eigenfunctions
and eigenvalues \cite{Risken96}: \begin{eqnarray}
P(\theta,z,t) & = & e^{\hat{L}(t-t_{0})}P_{\mathrm{initial}}(\theta,z)\nonumber \\
 & = & \sum_{k}a_{k}e^{\lambda_{k}(t-t_{0})}P_{k}(\theta,z)\label{eq:TimeEvolutionEF}\end{eqnarray}
with $P_{\mathrm{initial}}(\theta,z)=\sum_{k}a_{k}P_{k}(\theta,z)$.
Although the eigenfunctions of $\hat{L}$ form a basis, it is important
to note that they are not orthogonal. An important property is that
the mean value of all eigenfunctions except $P_{0}(\theta,z)$ is
zero:\begin{equation}
\int_{-\pi}^{\pi}d\theta\int_{-\infty}^{\infty}dz\, P_{k}(\theta,z)=0\end{equation}
To actually compute the spectrum and eigenfunctions we expand $P(\theta,z)$
into a set of complete orthonormal functions:\begin{equation}
P(\theta,z)=\sum_{m=0}^{\infty}a_{n,m}\psi_{n,m}(\theta,z)\label{eq:PExpansion}\end{equation}
with \begin{eqnarray}
\psi_{n,m}(\theta,z) & = & \left(2^{m+1}\sqrt{\pi\tau/2\tau_{c}}m!\right)^{-1/2}\nonumber \\
 &  & e^{in\theta}H_{m}\left(\sqrt{2\tau_{c}/\tau}z\right)e^{-z^{2}\tau_{c}/\tau}.\label{eq:BasisPsinm}\end{eqnarray}
This expansion obeys the imposed boundary conditions. In the $\theta$-directions
it consists of plane waves, while in the $z$-direction harmonic-oscillator
functions are used \cite{Gardiner85} with the Hermite polynomials
$H_{m}(z)$ \cite{Abramowitz72}. We now insert Eq.~(\ref{eq:PExpansion})
in Eq. (\ref{eq:FokkerPlanckCorrelated}). Multiplying from left with
$\psi_{n',m'}^{*}(\theta,z)$ and integrating over the whole domain
leads to a matrix eigenvalue equation for the $\left(a_{n,m}\right)$:\begin{eqnarray}
\lambda a_{n,m} & = & \left(-i\tau^{-1}(1+I_{0})n-\tau_{c}^{-1}m\right)a_{n,m}\nonumber \\
 &  & +(2\tau)^{-1}i(1-I_{0})n\,\left(a_{n-1,m}+a_{n+1,m}\right)\nonumber \\
 &  & -\frac{in\sigma}{2\sqrt{\tau\tau_{c}}}\left(\vphantom{\frac{1}{2}}(m+1)a_{n,m+1}+m\, a_{n,m-1}\right.\nonumber \\
 &  & \quad+\frac{1}{2}(m+1)\left(a_{n-1,m+1}+a_{n+1,m+1}\right)\nonumber \\
 &  & \quad\left.+\frac{1}{2}(m+1)\left(a_{n-1,m-1}+a_{n+1,m-1}\right)\right)\nonumber \\
 &  & +\left(\sqrt{2\tau\tau_{c}}\right)^{-1}\sqrt{(m+1)(m+2)}\, a_{n,m+2}\nonumber \\
 &  & -i\tau^{-1}\alpha\left(a_{n,m}c_{0}+\frac{1}{2}\sum_{k=1}^{K}\left(ic_{k}+s_{k}\right)a_{n-k,m}\right.\nonumber \\
 &  & \quad\left.+\left(iC_{k}-S_{k}\right)a_{n+k,m}\vphantom{\sum_{k=1}^{K}}\right)\nonumber \\
 & = & \sum_{n',m'}L_{n,m;n',m'}a_{n',m'}.\label{eq:MatrixEqCorrelated}\end{eqnarray}
 The coefficients $C_{k}$and $S_{k}$ denote the Fourier components
of $(1+\cos\theta)\tanh\left(\beta\tan\left(\theta/2\right)\right)$
of the expansion in $\cos\left(k\theta\right)$ and $\sin\left(k\theta\right)$
up to order $K$. To solve this eigenvalue problem numerically we
have to restrict the indices $n$ and $m$ to \begin{equation}
n\in\left\{ -N\ldots N\right\} ,\qquad m\in\left\{ 0\ldots M\right\} \end{equation}
Since the stationary density is very peaked for realistic firing rates,
we need many plane wave basis functions, i.e.~up to $N\approx10^{4}$.
With $M=50$ the matrix that we have to diagonalize will be of size
$10^{6}\times10^{6}$. To only represent this matrix in full form
would require $3.8\cdot10^{3}\textrm{GB}$ of storage capacity. We
note, however, that the matrix $\mathbf{L}$ in Eq.~(\ref{eq:MatrixEqCorrelated})
is very sparse, for $\alpha=0$ it connects an element $a_{n,m}$
even only to the elements $a_{n\pm1,m\pm1}$and $a_{n,m+2}$. For
$\alpha>0$ the number of nonzero entries in $\mathbf{L}$ solely
depends on the number of Fourier components $K$ of the AP onset term
of the generalized model. In general, however, the number of elements
in the matrix $\mathbf{L}$ is only of order $N\times M$, i.e.~very
sparse compared to its full size $N^{2}\times M^{2}$. This makes
it possible to use a high performance iterative algorithm, the Arnoldi-method
\cite{Trefethen97,Lehoucq98} to solve this eigenvalue problem numerically.
The time-dependent firing rate $\nu(t)$ is calculated using Eq.~(\ref{eq:LROutputRate}).

\section{Results}

\subsection{High frequency limit}

\subsubsection{Dynamics insensitive at action potential ($\theta_{s}=\pi$)}

For both types of input modulations the modulus of the right hand
side of Eq.~(\ref{eq:FirstOrderHighFreqLimit}) vanishes at $\theta=\pi$.
Therefore the $\tilde{P}(\theta,z,t)$ has to be at least of order
$\omega^{-2}$, such that the left hand side vanishes for $\omega\to\infty$.
Differentiation of Eq.~(\ref{eq:LinResFPG}) and subsequent reinsertion
leads to:\begin{equation}
\left(\omega^{2}+\hat{L}_{0}\right)\tilde{P}(\theta,z,t)=-\hat{L}_{0}\hat{L}_{1}P_{0}(\theta,z)\label{eq:SecOrderHighFreqLimit}\end{equation}
The right hand side does not vanish at $\theta=\pi$ in the case of
a mean current modulation and in the case of a modulation in the noise
amplitude. Since both sides have to be real valued, the modulus of
$\tilde{P}(\theta,z)$ has to be $\propto\omega^{-2}$ and the phase
$\varphi(\omega)$ goes to $-\pi$.

In the limit $\tau_{c}\to0$, i.e.~an uncorrelated input current,
the same argument holds in the case of a mean current modulation.
For a modulation in the noise amplitude, the right hand side of Eq.~(\ref{eq:SecOrderHighFreqLimit})
is zero, resulting even in a $\omega^{-3}$ decay and a phase lag
of $3\pi/2$.

\subsubsection{Generic case ($\theta_{s}\not=\pi)$}

For $\theta_{s}=\pi-\delta$, $\delta>0$ the right hand side of Eq.~(\ref{eq:FirstOrderHighFreqLimit})
does not vanish. This means that for large frequencies the rate modulation
$\nu_{1}(\omega)$ decays as $\omega^{-1}$ and the relative phase
shift $\varphi(\omega)$ is $-\pi/2$, which is the same asymptotic
decay as in conductance based neuron models. Table \ref{cap:Table:HFSummary}
summarizes the high frequency behavior of the generalized $\theta$-neuron
and compares it to the high-frequency limit of conductance based model
neurons and the classical leaky integrate-and-fire model. %
\begin{table}
\begin{center}\begin{tabular}{|c||c|c|c|c|}
\hline 
&
\multicolumn{4}{c|}{$\theta$-neuron}\tabularnewline
\hline
\hline 
&
\multicolumn{2}{c|}{$\theta_{s}=\pi$}&
\multicolumn{2}{c|}{$\theta_{s}\not=\pi$}\tabularnewline
\hline 
Noise correlation&
$\tau_{c}>0$&
$\tau_{c}\to0$&
$\tau_{c}>0$&
\multicolumn{1}{c|}{$\tau_{c}\to0$}\tabularnewline
\hline
\hline 
Mean modulation&
$\omega^{-2}$&
$\omega^{-2}$&
$\omega^{-1}$&
$\omega^{-1}$\tabularnewline
\hline
Noise modulation&
$\omega^{-2}$&
$\omega^{-3}$&
$\omega^{-1}$&
$\omega^{-1}$\tabularnewline
\hline
\hline 
&
\multicolumn{2}{c|}{LIF model}&
\multicolumn{2}{c|}{CB models}\tabularnewline
\hline
\hline 
&
$\tau_{c}>0$&
$\tau_{c}\to0$&
$\tau_{c}>0$&
$\tau_{c}\to0$\tabularnewline
\hline
Mean modulation&
$\omega^{0}$&
$\omega^{-1/2}$&
$\omega^{-1}$&
$\omega^{-1}$\tabularnewline
\hline
Noise modulation&
$\omega^{0}$&
$\omega^{0}$&
$\omega^{-1}$&
$\omega^{-1}$\tabularnewline
\hline
\end{tabular}\end{center}

\caption{\label{cap:Table:HFSummary}High frequency behavior of the generalized
$\theta$-neuron, the leaky integrate-and-fire model \cite{Brunel01,Lindner02}
and conductance based models. The response of a conductance based
model for $\tau_{c}>0$ and a mean current modulation was studied
in \cite{Fourcaud03}. The asymptotic response of the conductance
based model in the other cases follows from the same argument as for
the asymptotic response of the $\theta$-neuron and was confirmed
by direct numerical simulations (data not shown)}
\end{table}

We would like to point out, that the $\omega^{-2}$ and $\omega^{-3}$
decay of the classical $\theta$-neuron is only due to (i) the insensitivity
of the dynamics to inputs at $\theta=\pi$ and the symmetric up- and
downstroke of the action potential around $\theta_{s}=\pi$. Here,
both conditions are lifted by defining the spike phase at a different
value than $\pi$. Another way to induce a $\omega^{-1}$-decay would
be to change the right hand side of Eq.~(\ref{eq:GeneralizedThetaNeuron}),
such that $\hat{L}_{1}P_{0}$ does not vanish at $\theta=\pi$, e.g.
by introducing high order terms in $\cos\theta$. This would however
require a structural change of the oscillator dynamics. A second important
point to note is the independence of the high-frequency limit from
the dynamics at the action potential onset.

\subsection{Linear response transmission Functions}

\begin{figure*}
\begin{center}\includegraphics[%
  width=1.0\textwidth]{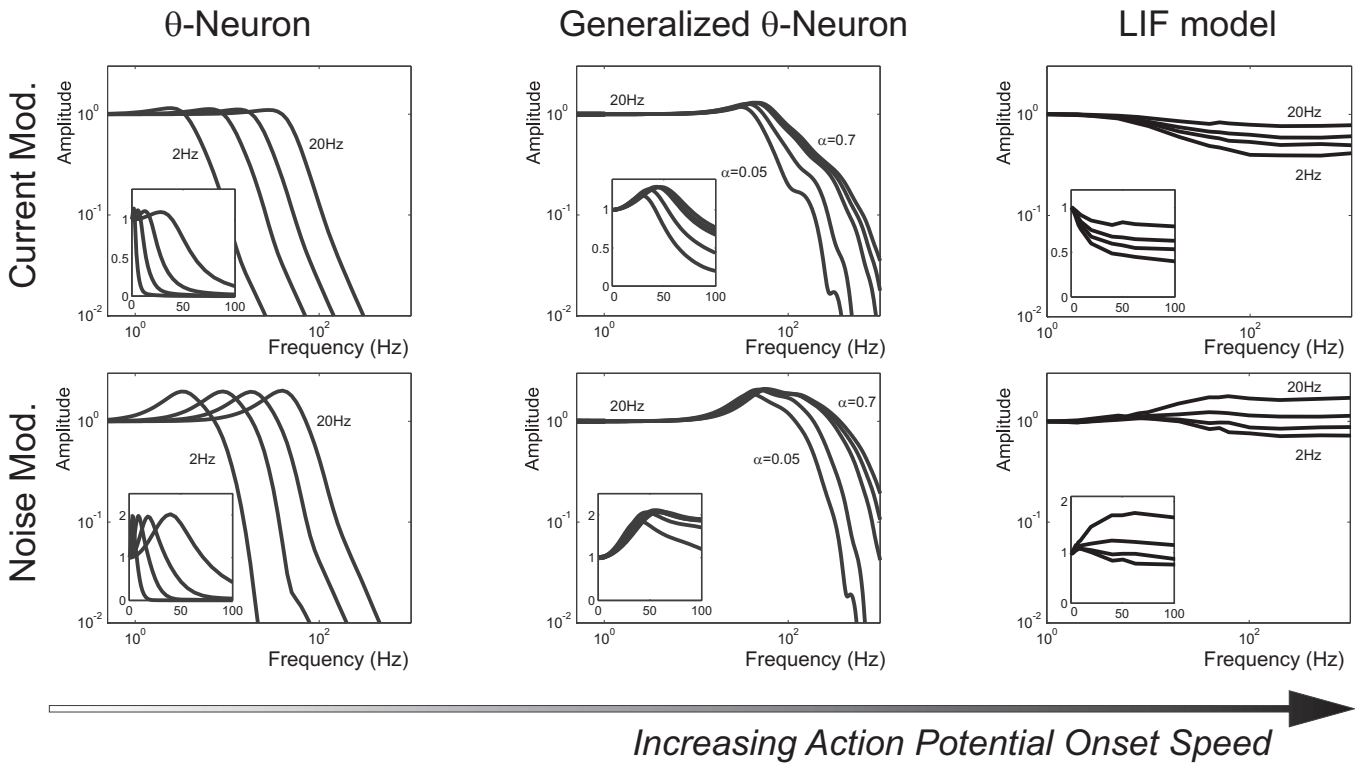}\end{center}

\caption{\label{cap:DynRespRes}Response amplitude for increasing values of
the action potential onset speed. In the left column the response
of the $\theta$-neuron for modulations in the mean input current
and the noise amplitude is shown for different values of the stationary
firing rate. The response exhibits a cut-off behavior, frequencies
larger than the stationary firing rate are strongly damped. The middle
column shows the response of the generalized model for both types
of modulation and a stationary rate of 20Hz. For increasing values
of the action potential onset speed the response amplitude grows for
frequencies in the interval from $100\textrm{Hz}$ to $1\textrm{kHz}$,
while the resonance maximum only slightly shifts to larger frequencies.
The response of the noise modulation is much larger in this interval
than the response to modulations in the mean current. For comparison
the right column shows the response of the leaky integrate-and-fire
(LIF) model. The response amplitude does not decay for large frequencies,
for modulations of the noise amplitude it can even grow with increasing
input frequencies. Parameters in the LIF simulation are as in \cite{Brunel01},
except $\tau_{s}=10\textrm{ms}$, $\sigma=5\textrm{mV}$, $I_{0}=14.6;16.2;17.5;19.5\textrm{mV}$
for a mean firing rate of $2,5,10,20\textrm{Hz}$.}
\end{figure*}
Using the matrix method described above, we computed the linear responses
to modulations in the mean input current and to modulations in the
noise amplitude. Figure \ref{cap:DynRespRes} summarizes the response
amplitude curves for the $\theta$-neuron model, the generalized $\theta$-neuron
model and compares them to direct numerical simulations of the response
of the leaky integrate-and-fire (LIF) model. 

The $\theta$-neuron exhibits a cut-off behavior in its response amplitude
to both types of input modulations. Frequencies up to the stationary
firing rate can be transmitted unattenuated larger frequencies are
strongly damped. For an increasing onset speed and fixed stationary
rate the resonance maximum shifts only to slightly larger frequencies,
a dramatic change, however, occurs at intermediate frequencies up
to 1kHz. In this regime the response amplitude is substantially lifted
to much larger transmission amplitudes. This effect is much more pronounced
for the case of a modulation in the noise amplitude than for modulations
in the mean input current, leading to an undamped response for frequencies
up to 200Hz. The LIF model, on the other hand, shows a completely
artificial response behavior. The transmission function, for both
types of modulations does not decay at all, even for frequencies up
to 1kHz. For modulations in the noise amplitude, the transmission
function can even grow for increasing stimulation frequencies.

\subsection{Nonlinear response for large stimulation amplitudes}

\begin{figure}
\begin{center}\includegraphics[%
  width=0.95\columnwidth]{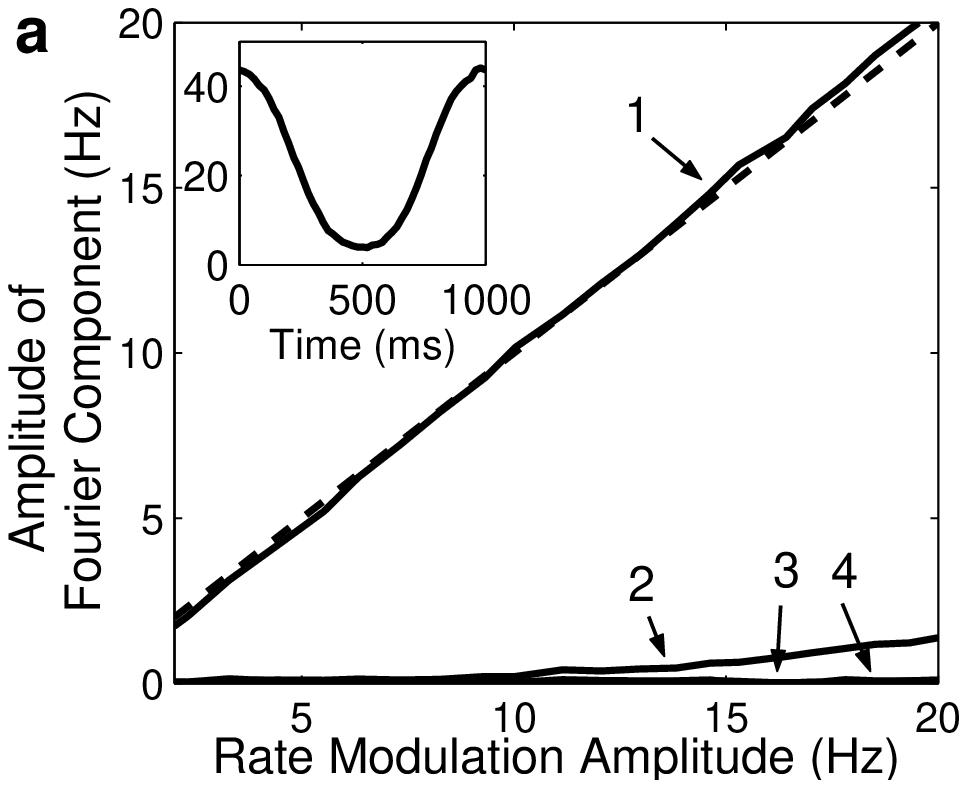}\hfill{}\includegraphics[%
  width=0.95\columnwidth]{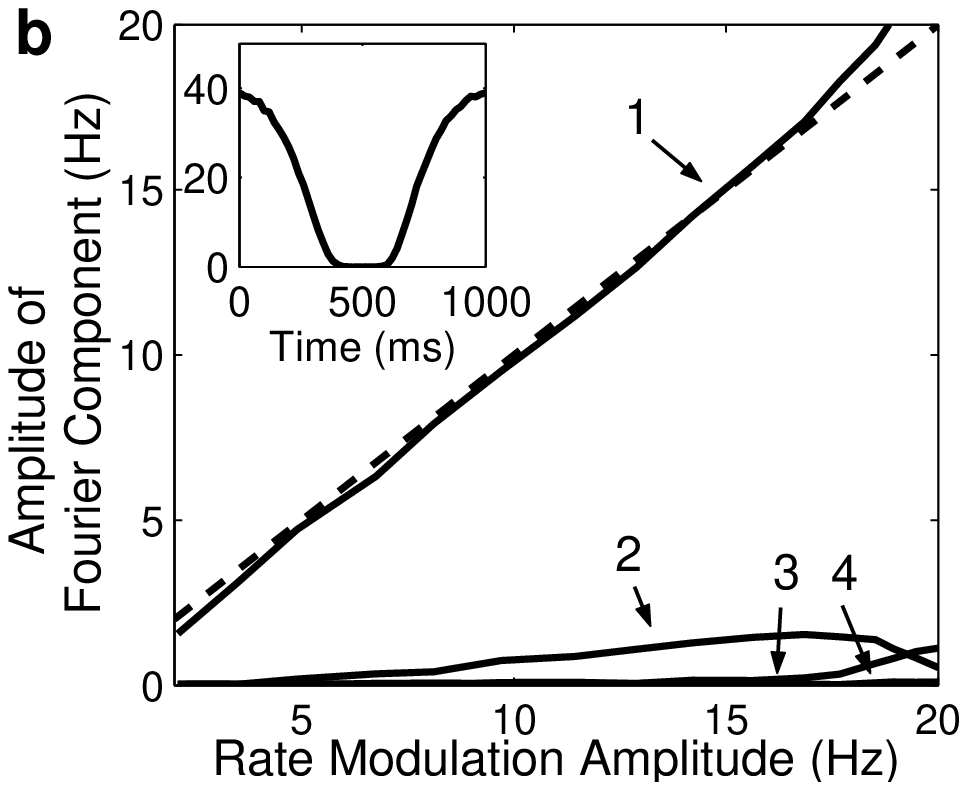}\end{center}

\caption{\label{cap:NonlinResp}Amplitude of the first four Fourier components
as a function of overall modulation amplitude of the population averaged
firing rate for (a) modulations in the mean input current and (b)
for modulations in the noise amplitude ($\alpha=0.7$). The mean output
rate is 20Hz, the modulation frequency 1Hz. The dashed line is the
diagonal. Up to amplitudes close to the mean output rate, the first
Fourier component is indistinguishable from the diagonal, indicating
that the response is essentially linear. Starting from amplitudes
comparable to the mean rate, the influence of higher order Fourier
components becomes substantial. The insets show the rate modulation
for a modulation amplitude of 20Hz.}
\end{figure}
So far we have only considered the linear response transmission function,
which is strictly speaking only valid in the limit in which the stimulation
amplitude goes to zero. Here we show, however, that the linear response
covers a large range of input amplitudes. In principle, we could use
the same matrix method employed for the linear response theory, taking
into account higher order Floquet modes \cite{Reichl88}. Here we
explore this regime, however, by direct numerical simulation of Eq.~(\ref{eq:GeneralizedThetaNeuron}).
Figure \ref{cap:NonlinResp} shows the amplitude of the first four
Fourier modes of the rate response as a function of the overall amplitude
of the rate modulation. For both types of modulations, the first Fourier
component clearly dominates the response up to amplitudes close to
the mean rate, where nonlinearities are naturally expected, as there
are no negative firing rates. This demonstrates that the linear response
theory, although rigorously valid for small modulation amplitudes
only, predicts the response in a large dynamical range of input amplitudes.

\section{Summary and Discussion}

The dynamical response properties of the generalized $\theta$-neuron
with adjustable AP onset speed were calculated in the presence of
a fluctuating correlated background noise. Methodologically we introduced
a new approach which is based on the expansion of the corresponding
Fokker-Planck operator to a complete set of orthonormal functions,
leading to a sparse matrix representation. We then computed the eigenvalues
and eigenfunctions of this matrix using an iterative scheme, the Arnoldi
method. The high frequency limit was calculated analytically. It turned
out, that the response amplitude decays as $\omega^{-\gamma}$, where
$\gamma$ depends on the kind of stimulation and, surprisingly, the
phase at which a spike is emitted. As soon as this point differs from
$\pi$, where the dynamics is insensitive to external inputs, the
exponent $\gamma$ is $1$, giving the same asymptotic response behavior
as conductance based neuron models. Using the eigenvalues and eigenfunctions
we then presented a method to evaluate the dynamic response to small
time-varying inputs. There we found that for the classical $\theta$-neuron
model the response exhibits a cut-off behavior: For a modulation in
the mean input current as well as for a modulation in the noise amplitude
frequencies above the stationary rate of the neuron were strongly
damped. In the generalized $\theta$-neuron the damping in the regime
up to 1kHz is substantially reduced for both types of input modulations
when the AP onset speed is increased, although the high frequency
limit is the same as in the classical $\theta$-neuron. The response
amplitude for the noise amplitude modulation is typically much larger
than the response amplitude for the mean input modulation. The linear
response theory, although only derived for small modulations of the
input current turned out to be valid in a large dynamical range, which
we demonstrated by direct numerical simulations. Amplitudes of the
rate modulation up to the mean output rate turned out to be well described
by the linear theory. 

Simple phenomenological, yet dynamically realistic models of cortical
neurons are of key importance for studies in theoretical neuroscience,
starting from studies on spike timing to large scale network simulations
or analytical network studies. While \emph{stationary} response properties,
such as mean firing rates or processes have been studied in many models,
which operate on long time scales, e.g.~adaptation (see e.g.~\cite{Benda03,Senn03,Izhikevich04}),
studies on the \emph{dynamic} response properties are rare. Most of
these studies consider the dynamic response in the class of integrate-and-fire
(IF) models \cite{Knight72,Brunel01,Lindner02,Fourcaud02}. In these
studies, it was demonstrated that IF models can relay incoming stimuli
instantaneously. Recently it was shown, however, that this response
behavior strongly disagrees with the response of conductance based
models and rather represents an oversimplification of the model than
a feature of real neurons \cite{Naundorf03,Fourcaud03}. While in
\cite{Naundorf03} the response properties of the classical $\theta$-neuron
were investigated, the authors of \cite{Fourcaud03} studied another
phenomenological neuron model, the EIF model, which mimics the dynamical
response properties of a conductance based model. Our study corroborates
and extends some of their results using a generalized model of the
classical $\theta$-neuron \cite{Kopell86,Gutkin98}, a canonical
model of conductance based neuron models, which exhibit type-I excitability
and which, in contrast to IF models, incorporates a dynamic action
potential onset. While the classical $\theta$-neuron model was originally
studied in the super-threshold, noise-free case\cite{Ermentrout84,Kopell86},
recent studies focused on the response in the presence of fluctuating
input currents \cite{Gutkin98,Lindner03,Brunel03}. These studies
indicated that in a large parameter regime the $\theta$-neuron exhibits
the same stationary response properties as cortical neurons, e.g.~a
realistic shape of the f-I curve and irregular firing in the subthreshold
regime. 

Despite these results, a major point of criticism questioning the
biological relevance of the model, remained: While the $\theta$-neuron
reflects the dynamics at the onset to repetitive firing, it lacks
the sharp action potential upstroke found in more detailed conductance
based models and real neurons \cite{Fourcaud03}. It was further argued
that this deficiency results in a high frequency limit of the linear
response amplitude, which decays too fast $\propto\omega^{-2}$, while
the linear response amplitude in conductance neuron models only decays
$\propto\omega^{-1}$. To address these issues we generalized the
classical $\theta$-neuron, incorporating an adjustable AP onset speed,
thereby mimicking the fast sodium activation at the action potential
onset. Surprisingly, our study reveals that the high frequency limit,
does not depend at all on the speed at the AP onset, but rather on
the phase variable, at which action potentials are emitted. If at
this point the dynamics is insensitive to external inputs, as in the
classical $\theta$-neuron, the decay of the linear response amplitude
is at least $\propto\omega^{-2}$, whereas the decay is always $\propto\omega^{-1}$
if the dynamics is not completely insensitive to external inputs,
as is the case in conductance based neuron models. Moreover, the full
transmission function reveals that the onset of the high-frequency
limit can be shifted to very high frequencies if the speed of the
AP onset is increased. These results question the relevance of the
high frequency limit as a criterion for the typical transmission speed
of neuron models.

For the computation of the linear response amplitude we did not resort
to direct numerical simulations, but used a method based on the eigenfunctions
and eigenvalues of the Fokker-Planck operator, describing the dynamics
of the probability density function in the presence of a temporally
correlated fluctuating input current. While this approach is in general
well-known (see e.g.~\cite{Risken96} and \cite{Knight00,Mattia02}
for an application to the non-leaky integrate-and-fire model in the
presence of an uncorrelated background noise), we derived a sparse
matrix representation, for which we computed eigenvalues and eigenfunctions
with very high numerical accuracy using a fast iterative scheme, the
Arnoldi method \cite{Trefethen97,Lehoucq98}. Compared to previous
studies on dynamical responses \cite{Brunel01,Fourcaud02,Fourcaud03},
this allowed for the computation of the linear response properties
with an accuracy that would be hard to meet by a direct simulation
of the single neuron dynamics.

Besides this, our results provide a direct link to experiments. In
a recent study \cite{Brumberg02} it was shown that the AP width in
neocortical neurons is strongly correlated with the critical frequency
up to which a neuron can phase lock to sinusoidal input stimulations.
This is indeed the same result we found for the generalized $\theta$-neuron:
With increasing AP onset speed the response amplitude shifts to larger
frequencies, enabling the model to respond to frequencies much larger
than its own stationary rate. In a second experimental study it was
demonstrated that cortical neurons subject to fluctuating input currents
adapt their instantaneous firing much faster when stimulated with
a step input in the noise amplitude than with a step mean input current
\cite{Silberberg04}. This behavior is well reproduced by the generalized
$\theta$-neuron. For increasing values of the AP onset speed, the
response amplitude at high frequencies is one order of magnitude larger
for the stimulation in the noise amplitude, compared to the stimulation
in the mean input current. Both results strongly suggest that the
generalized $\theta$-neuron, despite its simplicity and analytic
tractability, captures well the essence of the AP generating mechanism
of multidimensional conductance based neuron models. Future experimental
studies will have to show to what extent the generalized $\theta$-neuron
predicts the dependence of the dynamical response properties on the
AP generating mechanism.

\end{document}